\documentclass[aps,prl,twocolumn,superscriptaddress,showpacs]{revtex4-1}
\pdfoutput=1

\usepackage{graphicx}
\usepackage{color}
\usepackage{amssymb}

\begin{document}

\title{Local origin of global contact numbers in frictional ellipsoid packings}

\author{Fabian M. Schaller}
\email{fabian.schaller@physik.uni-erlangen.de}

\affiliation{Institut f\"ur Theoretische Physik, Friedrich-Alexander-Universit\"at Erlangen-N\"urnberg, 91058 Erlangen, Germany}
\affiliation{Max Planck Institute for Dynamics and Self-Organization (MPIDS), 37077 Goettingen, Germany}

\author{Max Neudecker}
\affiliation{Max Planck Institute for Dynamics and Self-Organization (MPIDS), 37077 Goettingen, Germany}

\author{Mohammad Saadatfar}
\affiliation{Applied Maths, RSPhysSE, The Australian National University, Australia}

\author{Gary W. Delaney}
\affiliation{CSIRO Mathematics, Informatics and Statistics, Clayton South, Victoria, Australia}

\author{Gerd E. Schr\"oder-Turk}
\email{gerd.schroeder-turk@fau.de}
\affiliation{Institut f\"ur Theoretische Physik, Friedrich-Alexander-Universit\"at Erlangen-N\"urnberg, 91058 Erlangen, Germany}

\author{Matthias Schr\"oter}
\email{matthias.schroeter@ds.mpg.de}
\affiliation{Max Planck Institute for Dynamics and Self-Organization (MPIDS), 37077 Goettingen, Germany}

\date{\today}

\begin{abstract}
In particulate soft matter systems the average number of contacts $Z$ of a  particle is an 
important predictor of the mechanical properties of the system. 
Using X-ray tomography, we analyze
packings of frictional, oblate ellipsoids of various aspect ratios $\alpha$, prepared at different
global volume fractions $\phi_g$. 
We find that $Z$ is a monotonously increasing function of $\phi_g$ for all $\alpha$. 
We demonstrate that this functional dependence can be explained by a local analysis where each particle is described  
by its local volume fraction $\phi_l$ computed from a Voronoi tessellation. 
$Z$ can be expressed as an integral over all values of $\phi_l$:  
$Z(\phi_g, \alpha, X) = \int   Z_l (\phi_l, \alpha, X) \; P(\phi_l | \phi_g) \;  d\phi_l$. 
The local contact number function $ Z_l (\phi_l, \alpha, X)$
describes the relevant physics in term of locally defined variables only, including possible higher order terms $X$.
The conditional probability $P(\phi_l | \phi_g)$
to find a specific value of $\phi_l$ given a global packing fraction  $\phi_g$ is found to be independent of $\alpha$ and $X$.
Our results demonstrate that for frictional particles a local approach is not only a theoretical requirement but also feasible. 
\end{abstract}

%45.70.-n    Granular systems
%45.70.Cc   Static sandpiles; granular compaction
%61.43.-j     Disordered solids
%81.70.Tx   Computed tomography
 
\maketitle

The average number of contacts $Z$ that a particle forms with its neighbors is 
the basic control parameter in the theory of particulate systems known as the 
jamming paradigm \cite{liu:10,vhecke:10} where $Z$ is a function of 
the difference between the global volume fraction $\phi_g$ and some critical value $\phi_c$.
For soft, frictionless spheres (a practical example would be an emulsion) 
this is indeed a good description \cite{katgert:10}
because additional contacts are formed by the globally isotropic compression of the particles which also increases $\phi_g$.
However, in frictional granular media such as sand, salt, or sugar the control of $\phi_g$ is not achieved by compression but by
changing the geometric structure of the sample; if we want to fill more grains into a storage container we do not compress   
them with a piston, but we tap the container a couple of times on the counter top. 

But if $Z$ and $\phi_g$  are not simultaneously controlled by a globally defined parameter such as pressure, the 
idea of a function  $Z(\phi_g)$ runs into an epistemological problem: 
contacts are formed at the scale of individual particles and their neighbors.
At this scale the global $\phi_g$ is not only undefined; it would even be impossible for a particle 
scale demon to compute $\phi_g$ by averaging over the volume of the neighboring particles.  
The spatial correlations between Voronoi volumes \cite{lechenault:06,zhao:12,zhao:13}
would require it to gather information from a significantly larger volume than the direct neighbors.

To date, only two theoretical approaches have studied $Z$ from a local perspective: 
Song {\it et al.}~\cite{song:08} used a mean-field ansatz to derive a functional dependence between $Z$ and 
the Voronoi volume of a sphere. This ansatz has recently been expanded to arbitrary shapes composed
of the unions and intersections of frictionless spheres \cite{baule:13,baule:14}.
Secondly, Clusel {\it et al.}~\cite{clusel:09,corwin:10} developed the granocentric model  which predicts the probability 
distribution of contacts in jammed, polydisperse emulsions. The applicability of the granocentric model to frictional 
discs has been shown in \cite{puckett:11}.  

The aim of this experimental study is to go beyond spheres and understand how the average $Z$ in packings of 
frictional ellipsoids originates from the local physics at the grain level. We find that, to a first approximation, 
the number of contacts an individual particle forms depends on only two parameters: 
the material parameter $\alpha$ which is the length ratio  between the short and the two (identical) long axes of the ellipsoids. 
And a parameter that characterizes the  cage formed by all the neighboring particles:
the local volume fraction $\phi_l$ which is the particle volume divided by the volume of its
Voronoi cell. 

Frictional ellipsoids used in experiments \cite{man:05,farhadi:14,xia:14,wegner:14}  exhibit a number of differences
to the frictionless ellipsoids often studied numerically 
\cite{donev:04a,donev:04b,mailman:09,zeravcic:09,schreck:10,mkhonta:13,stenzel:14}. 
The latter have been found to form packings with less
than the number of contacts required for isostaticity, which is defined as having enough 
constraints to block all degrees of freedom of the particles
\cite{donev:04a,schreck:10,baram:12}.  
This apparent paradox has been resolved by Donev {\it et al.}~\cite{donev:07}, 
who showed that in this analysis the contacts can not be treated as the contacts between frictionless spheres:
the curvature of the ellipsoids blocks rotational degrees of freedom even in the absence of friction.
In contrast, we find packings of frictional ellipsoids to be hyperstatic over the whole
range of $\phi_g$ studied, in agreement with numerical simulation including friction \cite{guises:09,delaney:11}. 

\begin{figure}[t]
    \begin{picture}(240,90)
        \put(0,5){\includegraphics[width=2cm]{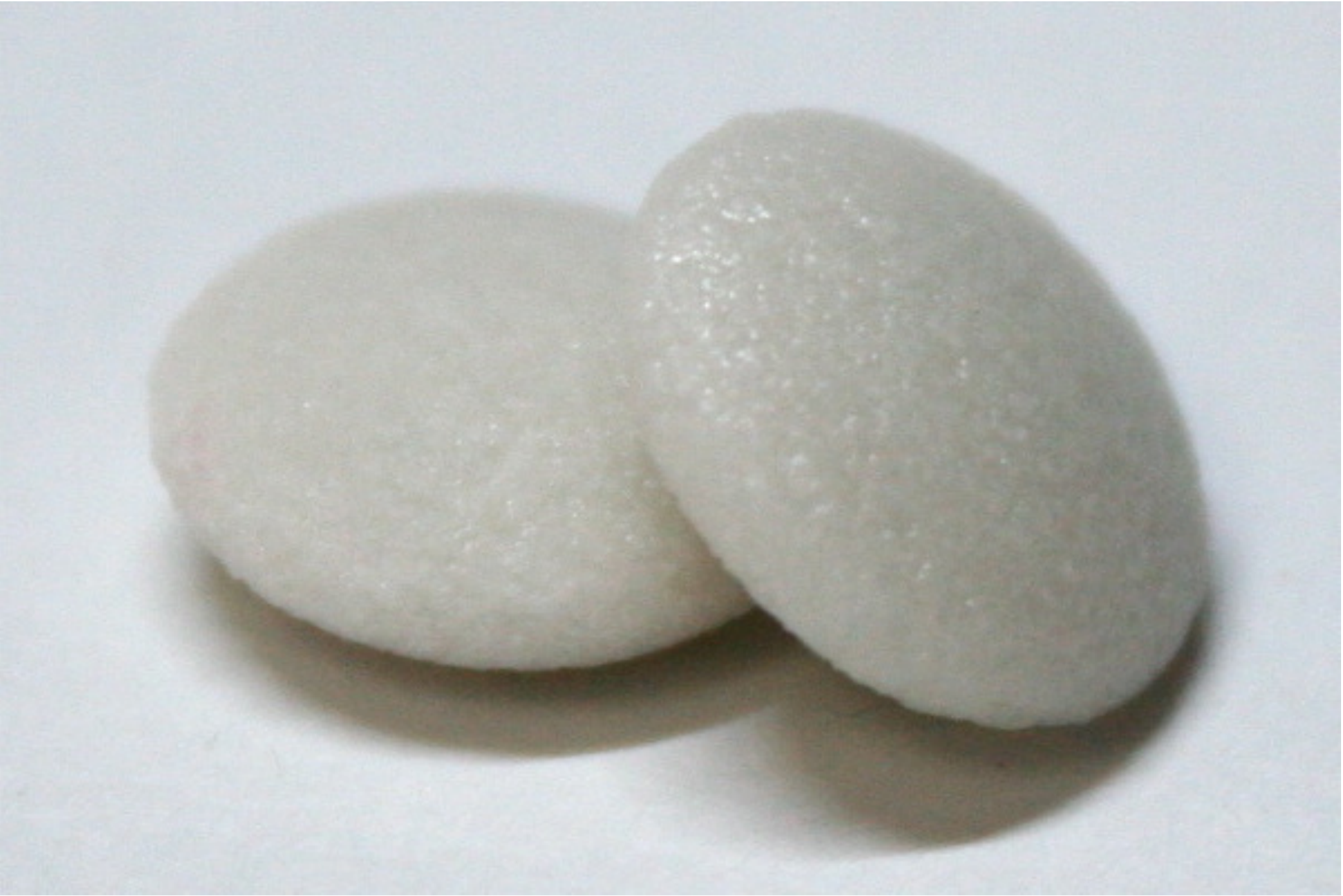}}
        \put(0,55){\includegraphics[width=2cm]{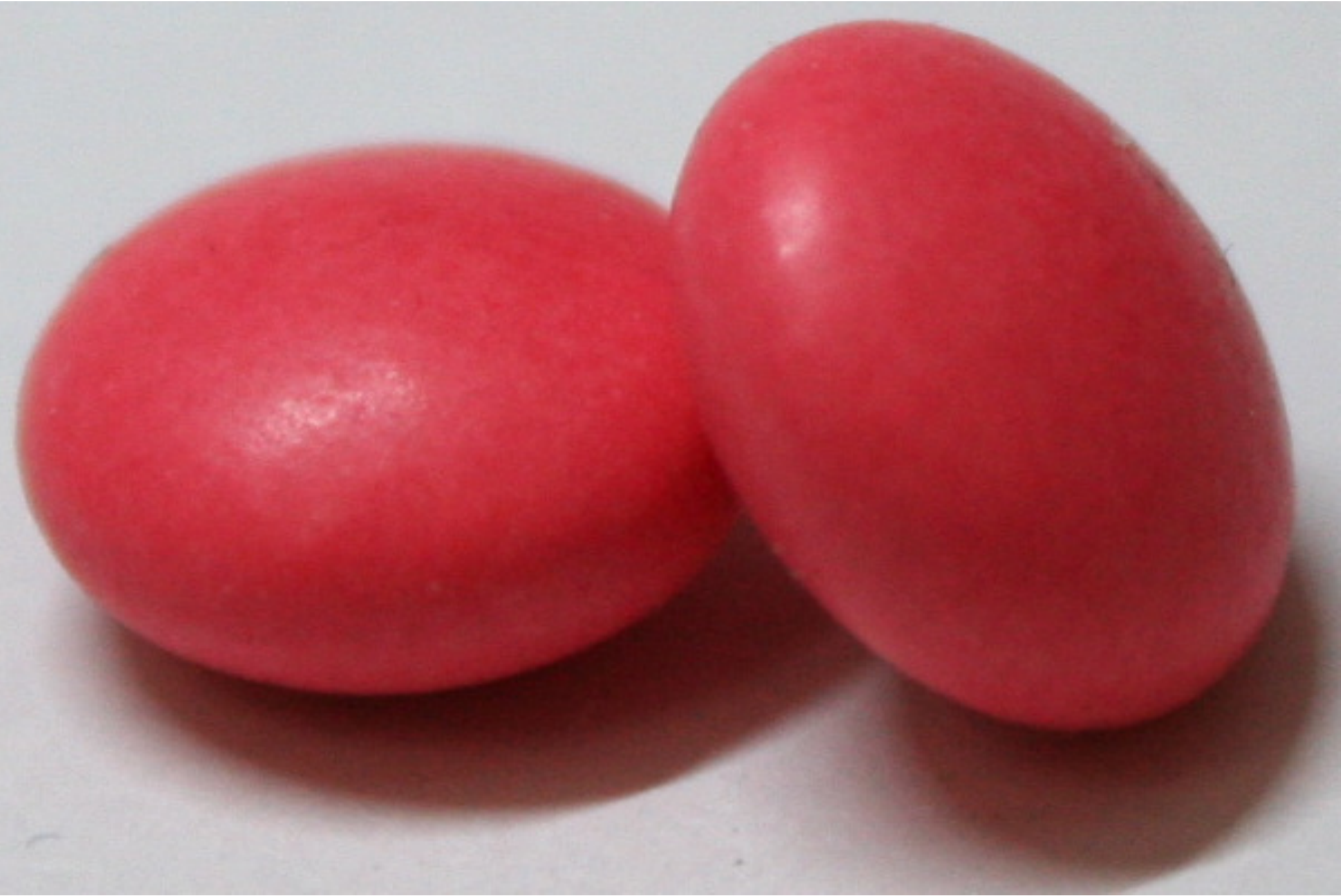}}
        \put(60,0){\includegraphics[width=3cm]{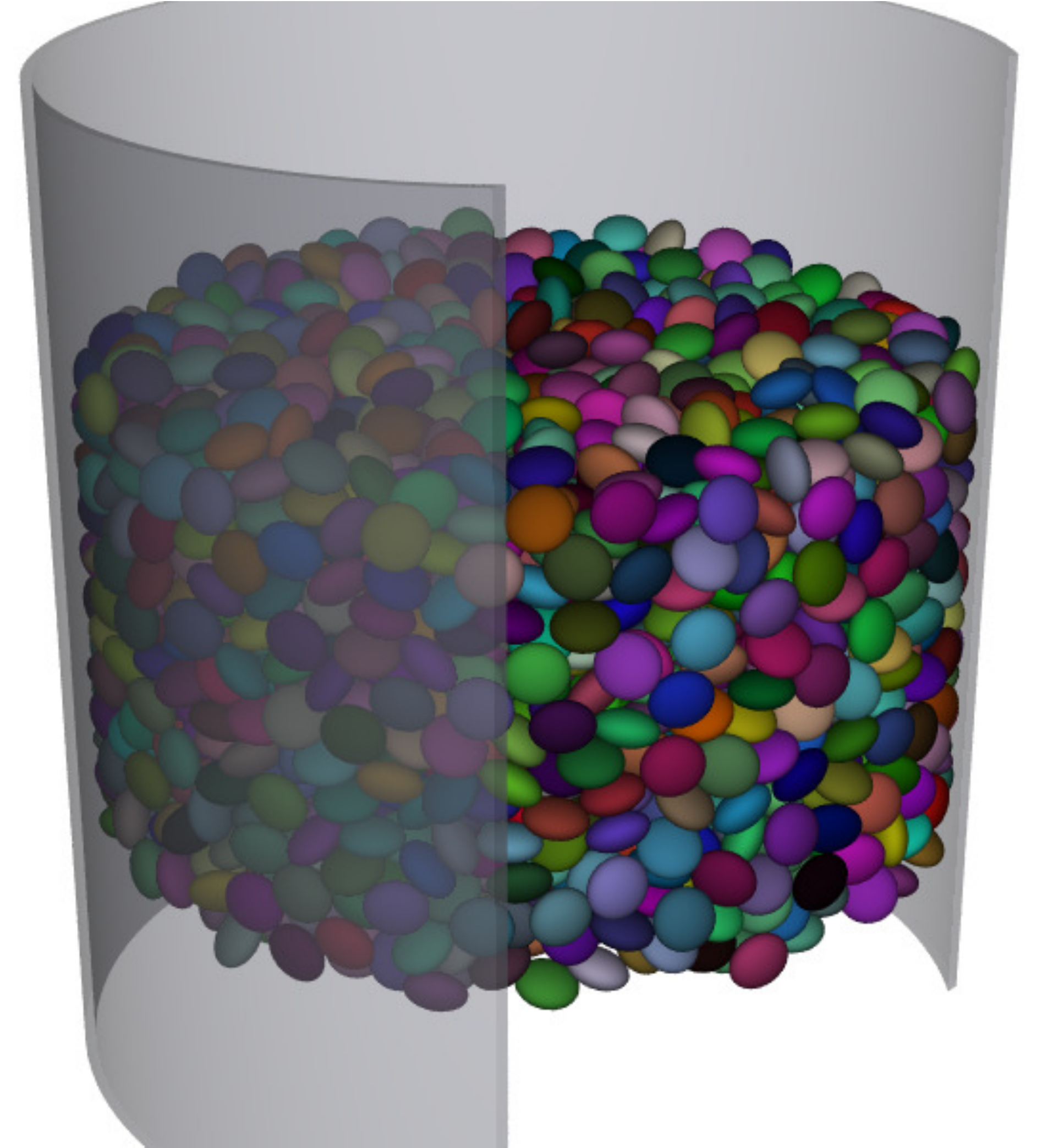}}
        \put(145,5){\includegraphics[width=3.3cm]{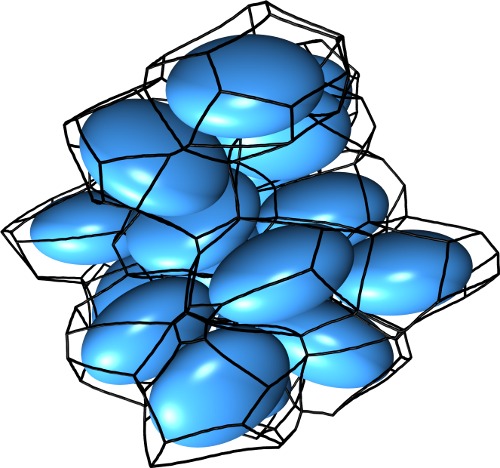}}
        \put(0,85) {\bf a)}
        \put(0,34) {\bf b)}
        \put(65,85){\bf c)}
        \put(150,85){\bf d)}
    \end{picture}

\caption{   a) Pharmaceutical placebo pills with $\alpha=0.59$.
            b) Gypsum particles made with a 3D printer with $\alpha =0.40$.
            c) Rendering of the particles detected in a X-ray tomogram.
            d) The black wire frame indicates the Voronoi cells of the ellipsoids.
\label{fig1}
    }
\end{figure}
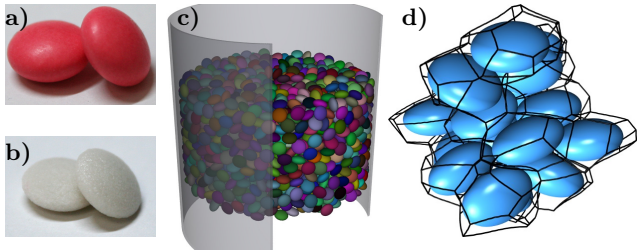

{\it Particles and preparation.-} We study two different types of oblate ellipsoids, 
the properties of which are summarized in table \ref{tab1}.
Figure \ref{fig1} a)  shows pharmaceutical placebo pills  (PPP) with $\alpha$ = 0.59 produced by Weimer Pharma GmbH. 
Due to their sugar coating, their surface is rather smooth; their static coefficient of friction $\mu_s$ against paper
is 0.38 (measured using a small sledge on a slowly raised inclined plane). 
The second particle type displayed in figure \ref{fig1} b) are gypsum ellipsoids cured with resin, produced  
with a 3D printer (Zprinter 650, Z corporation). The 
aspect ratio of these 3DP particles ranges from 0.4 to 1 (i.e.~spherical), their  
rougher surface results in values of $\mu_s$ between 0.67 and 0.75. Due to the production process, the 3DP particles
have hummocks of up to 100 $\mu$m on their short axis. As a consequence their volume deviates up to 3\% from a perfect 
ellipsoid, compared to 1\% for the PPP particles.

\begin{table*}
\begin{tabular}{c  c || c | c || c | c || c | c || c | c | c}
aspect  &                                                   & \multicolumn{2}{c||}{half axis}&type& friction       & particles&  number of & \multicolumn{3}{c}{Empirical fit parameters}   \\
ratio   &                                                   & short          & long        &     &coefficient      & in core  &  analyzed  & \multicolumn{3}{c}{for $Z_l (\phi_l, \alpha)$} \\ 
$\alpha$&                                                   & [mm]           & [mm]        &     &$\mu_s$          & region   &  packings  & $\ \quad a \ \quad$  & $\ \quad b \ \quad$   & $\ \quad c \;\quad$\\ \hline
spheres & \includegraphics[width=0.2cm]{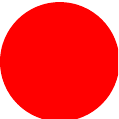}      & 3.1            &             & 3DP & $0.75 \pm 0.07$ & 660-850  &  15        & 60.4 & -52.2 & 14.8\\
0.80    & \includegraphics[width=0.2cm]{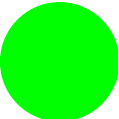}      & 2.65           & 3.30        & 3DP &$ 0.75 \pm 0.05$ & 750-850  &  17        & 60.4 & -52.4 & 15.1\\
0.60    & \includegraphics[width=0.2cm]{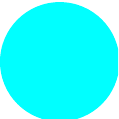}      & 2.20           & 3.75        & 3DP &$ 0.67 \pm 0.03$ & 620-710  &  16        & 44.7 & -31.0 & 8.4 \\
0.59    & \includegraphics[width=0.2cm]{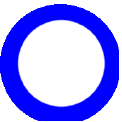} & 2.15           & 3.55        & PPP & $0.38 \pm 0.05$ & 850-910  &  15        & 63.5 & -53.7 & 15.4\\
0.40    & \includegraphics[width=0.2cm]{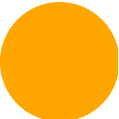}      & 1.60           & 4.00        & 3DP & $0.67 \pm 0.05$ & 620-730  &  10        & 25.3 & -10.7 & 3.9
\end{tabular}
\caption{Material properties of the particles. The first column displays the
color code used in figures \ref{fig:global_Z} to  \ref{fig:phi_l}.
Error-bars on $\mu_s$ are standard deviations over 15 experiments.
The last three columns show the empirical fit parameters for $Z_l (\phi_l, \alpha)$ according to equation \ref{eq:z_l_fit}.
\label{tab1}
    }
\end{table*}

Samples are prepared by first creating a loose packing of ellipsoids inside a 
plexiglass cylinder with an inner diameter of 104 mm; then the samples are tapped in order to increase $\phi_g$
to the desired value. We use three different protocols to prepare the initial loose samples, they are indicated 
by different symbols in the figures below. However, our results do not seem to depend on the initial 
preparation method, details of which can be found in the supplemental material.
Except for the loosest samples, the  packings are compactified by applying sinusoidally shaped pulses on
an electromagnetic shaker (LDS V555). The width of the pulses is 50 ms and the peak acceleration 2 $g$
(where $g=9.81 \mathrm{m/s^2}$). At a repetition rate of 3 Hz up to 1500 taps are applied to prepare the
highest values of $\phi_g$.

{\it Image analysis. --} Tomograms of the prepared
packings are acquired using X-ray computed tomography (GE Nanotom) with  a resolution of 64 $\mu$m per voxel.
The resulting three-dimensional gray scale image is the starting point for the identification of 
all particle centers and orientations (c.f.~figure \ref{fig1}c) using the methods described in \cite{schaller:13}. 
To reduce boundary effects,
only particles with centers that are at least two long axes away from the container walls were included in our analysis; 
table \ref{tab1} lists the numbers of these core particles. To assure 
spatial homogeneity, we discard all experiments where the standard deviation of the azimuthally averaged volume fraction 
is larger than 0.66\%.  Similarly, to exclude packings with a too large degree of local order we only  consider samples with
$\theta > 0.5$ rad where $\theta$ is the average angle of the short axis with respect to gravity
with $\theta = 1$ corresponding to a random orientation (see supplemental material). 
The particle positions and orientations of all experiments reported 
here can be downloaded from the Dryad repository \cite{dryad}.

From the geometrical representation of the sample we determine the average $Z$ using 
the contact number scaling method \cite{schaller:13}. Finally, the Voronoi cells 
of the particles are computed with the algorithm described in \cite{schaller_set:13}. 
Figure \ref{fig1} d) displays the Voronoi tessellation of a small subset of particles. 
By dividing the volume of the particle by the volume of the Voronoi cell we obtain for each particle its
local volume fraction $\phi_l$; the harmonic mean of all particles in the core region corresponds
to the global volume fraction $\phi_g$.

\begin{figure}
       \includegraphics[width=8cm]{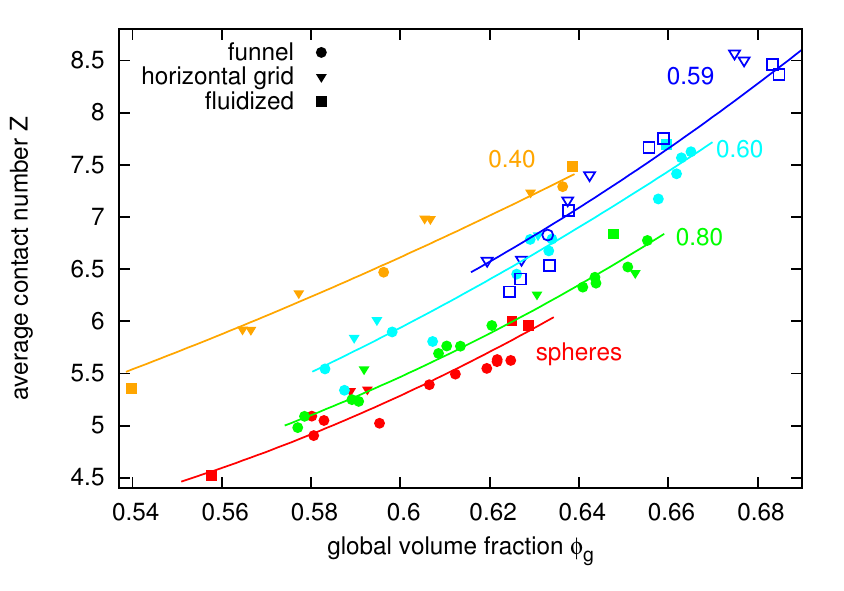}
    \caption{Contact number as a function of the global volume fraction.
Lines correspond to equation \ref{eq:ansatz_numint}, which  is  the numerical integration of the local theory presented here.
The different symbols indicate preparation of the initial packing, which is then compactified for all but the loosest samples by tapping.
The different aspect ratios and particle types are indicated by different colors, see table \ref{tab1}.
\label{fig:global_Z}
}
\end{figure}

{\it The average contact number} $Z$  as a function of $\phi_g$ is displayed in figure \ref{fig:global_Z}.
The main conclusion of figure \ref{fig:global_Z} is that the global average of $Z$ depends on both 
$\phi_g$ and  $\alpha$.
As expected for frictional particles \cite{silbert:02,zhang:05,shundyak:07,henkes:10,neudecker:13}, 
the contact number of all samples is significantly above the isostatic 
value of four \cite{note_isostatic}.

{\it Switching to a local ansatz. --} As discussed in the introduction, the formation of contacts between particles 
needs to be explained solely by parameters which are well defined on the particle level. We therefore start with an 
ansatz: 
\begin{equation}
    \label{eq:ansatz}
   Z (\phi_g, \alpha, X) = \int  Z_l(\phi_l, \alpha, X) \; P(\phi_l | \phi_g, \alpha, X) \;  d\phi_l
\end{equation}
Here the contact function $Z_l(\phi_l, \alpha, X)$ represents the local physics i.e.~the number of contacts formed 
by a particle of shape $\alpha$, inside a Voronoi cell of size $\phi_l$ and potentially characterized by   
further locally defined variables $X$ such as friction, fabric anisotropy, or 
measures of local order. 
$P(\phi_l | \phi_g, \alpha, X)$ is the conditional probability to find a particle with $\phi_l$ 
in a given packing; an integration over all values of $\phi_l$ will result in the global value of $Z$.

In order to measure how $Z_l$ depends on $\phi_l$, we determine the local contact number for each ellipsoid, 
see supplemental material.
Figure \ref{fig:phi_z_l} a) shows $Z_l(\phi_l)$ curves for all our experiments. The main point 
here is that in agreement with
our ansatz the curves for the 3DP particles do not depend on the global volume fraction $\phi_g$.
This result has been previously only shown for spheres \cite{aste:06}.
For the PPP particles the collapse is less conclusive, we discuss possible reasons below.
In consequence, we take for each value of $\alpha$ the average over all experiments, the resulting
$Z_l (\phi_l, \alpha)$ curves are shown in figure \ref{fig:phi_z_l} b). Here we have ignored not only $\phi_g$
but also all higher order terms $X$ because within the resolution of our experiments
we were not able to discern between different possible candidates. For a discussion of e.g.~$X$ 
being the orientation of the short axis see the supplemental material.
In order to obtain an phenomenological description for $Z_l$ we perform for each aspect ratio 
a parabola fit using: 
\begin{equation}
    \label{eq:z_l_fit}
Z_l (\phi_l, \alpha) = a \, \phi_l^2 + b \, \phi_l + c
\end{equation}
The results are displayed in figure \ref{fig:phi_z_l} b),
the values of the fit parameters $a$, $b$, and $c$ are listed in table \ref{tab1}.

Fitting equation  \ref{eq:z_l_fit} is a purely phenomenological approach, it is justified only by the absence 
of any theoretical predictions for frictional ellipsoids. The only analytical result available is for spheres
\cite{song:08}, it is in good agreement with our data (without any fit parameters)
as shown in the inset of figure \ref{fig:phi_z_l}. However, as discussed in the supplementary material, 
this result can not be easily generalized to frictional ellipsoids. Also included in the supplemental material
are fits which show that even a local re-interpretation of the jamming paradigm does fail to describe the physics.

\begin{figure}
       \includegraphics[width=8cm]{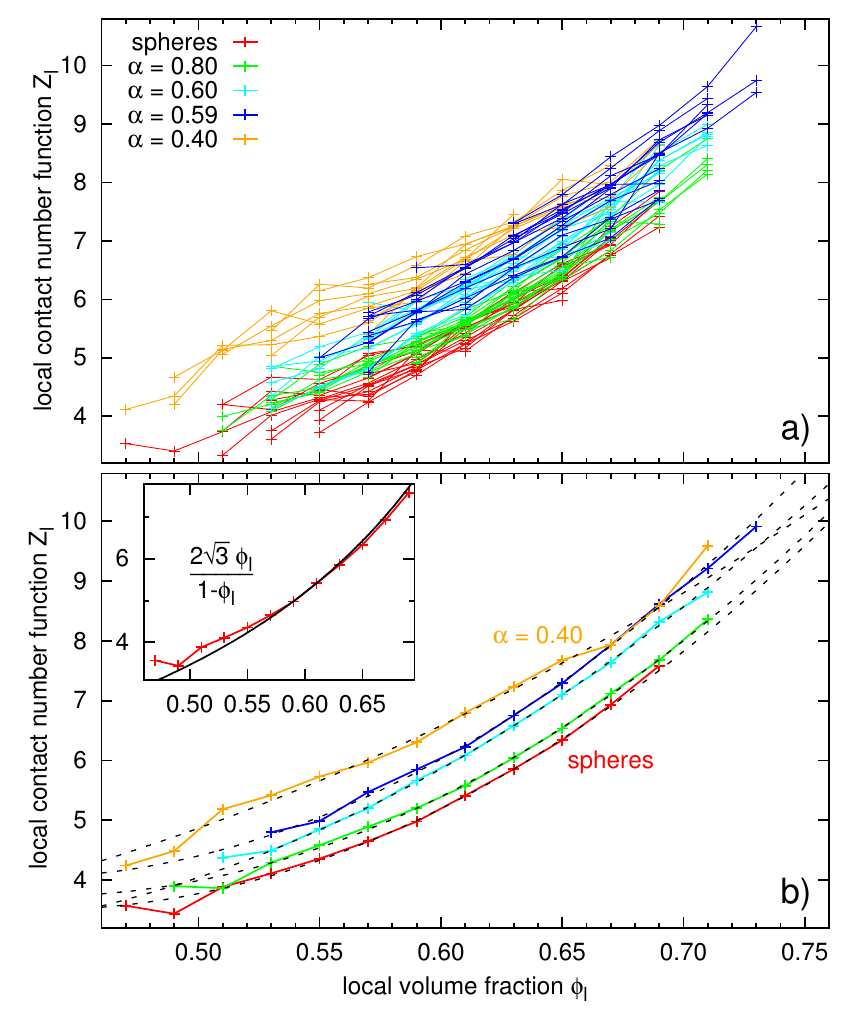}
\caption{Measuring the  local contact number function $Z_l$ which describes how many contacts an average particle with a 
local area fraction $\phi_l$ will form. 
In panel a) each line corresponds to a single experiment, i.e.~a single datapoint in 
figure  \ref{fig:global_Z}. Each cross represents the average number of contacts a particle with 
this value of $\phi_l$ (using a bin size of 0.02) will form.
The colored lines in panel b) are averages over all data sets (i.e.~different values of 
$\phi_g$) displayed in the upper panel. The black dashed lines are parabolic fits according to 
equation \ref{eq:z_l_fit}.
The inset shows the theoretical result from Song {\it et al.}~\cite{song:08} for spheres compared with our sphere data.
\label{fig:phi_z_l}
}
\end{figure}

{\it Properties of the local volume fraction distribution.--}
\begin{figure}
    \begin{picture}(235,290)
       \put(0,0){\includegraphics{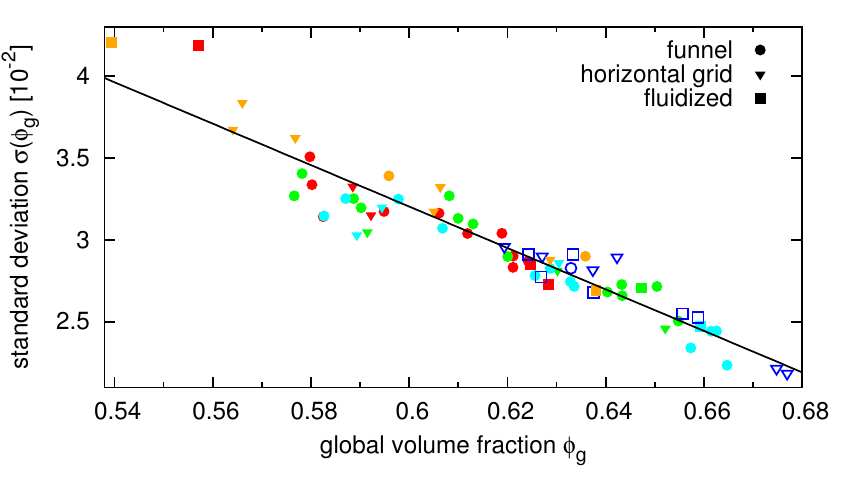}}
        \put(0,135){\includegraphics{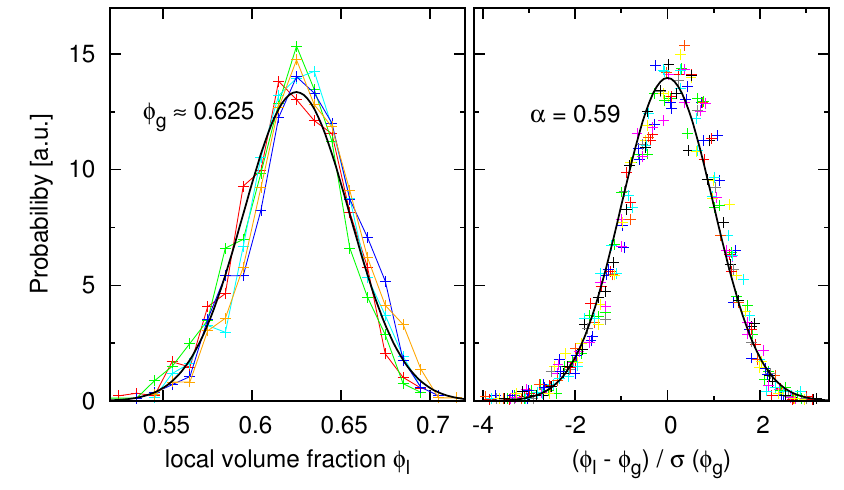}} 
     \put(40,255){\large \bf a)}
      \put(150,255){\large \bf b)}
        \put(40,35){\large \bf c)}
    \end{picture}
\caption{
Scaling properties of the local volume fraction distribution $P(\phi_l)$.
a) For a given global volume fraction (here $\phi_g \approx$ 0.625) the probability of finding a specific value 
of $\phi_l$ does not depend on $\alpha$. 
b) $P$ can be rescaled using $\phi_g$ and the standard deviation of the local volume fractions $\sigma$. Shown here 
are all experiments with $\alpha = 0.59$.
The black lines in panels a and b are Gaussian fits using equation  \ref{eq:Pfit}.
c) The standard deviations of the local volume fraction distribution depends only on $\phi_g$.
Panel includes all experiments shown in figure \ref{fig:global_Z}.
The black line is a linear fit resulting in equation  \ref{eq:sigma}.
\label{fig:phi_l}
    }
\end{figure}
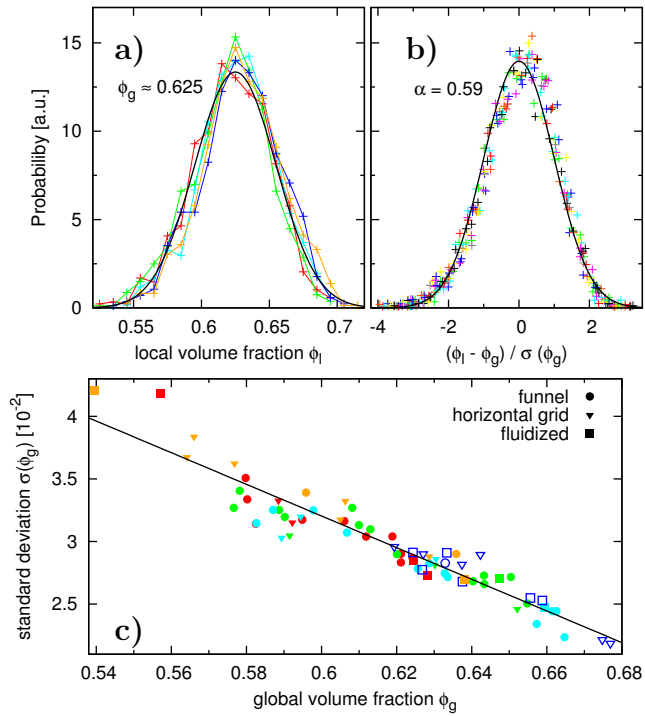
Figure \ref{fig:phi_l} reveals a number of interesting scaling properties of $P(\phi_l)$.
Panel \ref{fig:phi_l} a) shows $P(\phi_l)$ for all different aspect ratio at $\phi_g \approx$ 0.625. 
The good agreement indicates that $P(\phi_l)$ is independent of $\alpha$.
In figure \ref{fig:phi_l} b) a rescaled $P$ is plotted for all values of $\phi_g$. This demonstrates, that 
the mean (aka $\phi_g$) and the standard deviation of the local volume fraction distribution $\sigma({\phi_g})$ 
are sufficient to describe $P$.
This result has previously only been known for spheres \cite{aste:07,StarrGlotzer2002} and discs \cite{puckett:11}.
Finally, figure \ref{fig:phi_l} c) demonstrates that the standard deviation  $\sigma(\phi_l)$ of 
the local packing fraction distribution depends only on $\phi_g$ and not $\alpha$. 

Together, these results show that $P(\phi_l | \phi_g, \alpha, X)$  in equation \ref{eq:ansatz} can be replaced by 
 $P(\phi_l| \phi_g)$:
\begin{equation}
    \label{eq:ansatz_final}
    Z(\phi_g, \alpha, X) = \int  Z_l (\phi_l, \alpha, X) \; P(\phi_l | \phi_g) \;  d\phi_l    
\end{equation}
The advantage of this ansatz is a clear separation of the contact number problem into 
the local physics at the grain level and a probabilistic term connecting 
the local and the global volume fraction. Please note that without a better understanding of the origin of the scaling 
properties shown in figure \ref{fig:phi_l} it is not possible 
to decide on the causality between $\phi_l$ and  $\phi_g$. So writing  $P(\phi_l | \phi_g)$ can 
imply either that $\phi_g$ is the cause of the observed $P(\phi_l)$ or that $\phi_g$ can be seen 
to follow from the prepared $P(\phi_l)$.

In order to get an empirical expression for $Z$ we fit the local packing fraction distribution $P(\phi_l | \phi_g)$
with a Gaussian\footnote{We approximate $P(\phi_l | \phi_g)$ by a Gauss function, as this allows explicit integration of equation \ref{eq:z_l_fit} leading to \ref{eq:ansatz_numint}. Note the differences to the $k$-Gamma distributions \cite{aste:07} for the Voronoi cell volumes of spheres (that require the minimal possible Voronoi volume as an additional parameter, unknown for ellipsoids) are small.}
\begin{equation}
  \label{eq:Pfit}
    P(\phi_l | \phi_g) = \frac{1}{\sigma(\phi_g) \, \sqrt{2 \pi}} \; e^{\frac{-(\phi_l - \phi_g)^2}{2 \sigma(\phi_g)^2}}
\end{equation}
and the dependence of $\sigma$ on $\phi_g$ with a linear equation which yields: 
\begin{equation}
    \label{eq:sigma}
    \sigma(\phi_g) = - 0.126 \, \phi_g + 0.109
\end{equation}
Both fits are displayed as black lines in figure \ref{fig:phi_l}.
Entering equations \ref{eq:z_l_fit}  and  \ref{eq:Pfit}  into equation \ref{eq:ansatz_final}
and performing the integration leads to:
\begin{equation}
    \label{eq:ansatz_numint}
    Z(\phi_g, \alpha) = a \,\sigma(\phi_g)^2 + a \, \phi_g^2 + b \, \phi_g + c
\end{equation}
with $\sigma$ according to equation \ref{eq:sigma} 
and $a,b,c$ as shown in table \ref{tab1}.

A comparison of our experimental data with equation \ref{eq:ansatz_numint} is shown in 
figure \ref{fig:global_Z}. 
The good agreement for all 3DP particles 
demonstrates the validity of our ansatz equation \ref{eq:ansatz_final}.
For the PPP particles with $\alpha$ =0.59 the agreement is only fair, pointing to the need for an additional 
parameter $X$ in $Z_l(\phi_g, \alpha, X)$.
However, the experimental scatter does not allow us to assess the type of higher order corrections required.
The need for inclusion of such a parameter can also stem from the history-dependent behavior
of frictional particles. It has recently been shown for 
spheres \cite{agnolin:07} and tetrahedra \cite{neudecker:13,thyagu:15} 
that for identical $\phi_g$ the contact number can depend on the preparation history; 
modeling such behavior will require the addition of further locally defined parameters.

{\it Conclusion. --} The global contact numbers of packings of frictional spheres and ellipsoids can be 
explained by an ansatz which combines a local contact function and a conditional probability. 
The contact function does depend solely on parameters defined on the particles scale, 
including the local volume fraction and the aspect ratio of the particles.
The conditional probability to find a particle with a specific local volume fraction
is sufficiently described by the global volume fraction alone. 
We expect our results, available also as open data, to be a valuable reference point for the generalization
of existing theoretical approaches such as the granocentric model \cite{clusel:09,corwin:10} 
or the statistical mechanics approach to granular media \cite{song:08,baule:13} towards frictional granular matter. 
Extensions of our contact function including other locally defined parameters, such as e.g. the fabric anisotropy,
should be able to describe non-isotropic effects, such as observed in
shear-jammed frictional packings \cite{bi:11,grob:14}.

We thank Weimer Pharma GmbH for providing the PPP samples,
Martin Brinkmann, Stephan Herminghaus, Marco Mazza, Klaus Mecke and Jean-Fran\c{c}ois M\'etayer for valuable discussions   
and Matthias Hoffmann for programming support. 
We acknowledge funding by the German Science Foundation (DFG) through the research group
''Geometry and Physics of Spatial Random Systems'' under grant no SCHR-1148/3-1.


\begin{thebibliography}{44}
\bibitem [{not()}]{note_isostatic}%
  \BibitemOpen
  \href@noop {} {}\bibinfo {note} {Frictional contacts fix three force
  components, one normal and two tangential. These are shared between the two
  particles involved, so per particle each contact provides 1.5 constraints. As
  each particle posseses six degrees of freedom, at minimum four contacts are
  needed to block all of them.}\BibitemShut {Stop}%
\end{thebibliography}
\end{document}